# The PostScript files for the figures are appended at the end.

\documentstyle[twoside,12pt]{article}
\catcode`\@=11
\long\def\@makefntext#1{ 
\protect\noindent \hbox to 3.2pt {\hskip-.9pt
$^{{\eightrm\@thefnmark}}$\hfil}#1\hfill} 

\def\thefootnote{\fnsymbol{footnote}}
 \def\@makefnmark{\hbox to 0pt{$^{\@thefnmark}$\hss}}  

\def\ps@myheadings{\let\@mkboth\@gobbletwo
\def\@oddhead{\hbox{} 
\rightmark\hfil\eightrm\thepage}
\def\@oddfoot{}\def\@evenhead{\eightrm\thepage\hfil 
\leftmark\hbox{}}\def\@evenfoot{}
\def\sectionmark##1{}\def\subsectionmark##1{}}

\oddsidemargin=\evensidemargin
\renewcommand{\thefootnote}{\fnsymbol{footnote}}

\newcounter{sectionc}\newcounter{subsectionc}\newcounter{subsubsectionc}
\renewcommand{\section}[1] {\vspace{12pt}\addtocounter{sectionc}{1}
\setcounter{subsectionc}{0}\setcounter{subsubsectionc}{0}\noindent
	{\bf\thesectionc. #1}\par\vspace{5pt}}
\renewcommand{\subsection}[1] {\vspace{12pt}\addtocounter{subsectionc}{1}
	\setcounter{subsubsectionc}{0}\noindent
	{\bf\thesectionc.\thesubsectionc. {\kern1pt \bf\it #1}}\par\vspace{5pt}}
\renewcommand{\subsubsection}[1] {\vspace{12pt}\addtocounter{subsubsectionc}{1}
	\noindent{\thesectionc.\thesubsectionc.\thesubsubsectionc.
	{\kern1pt \it #1}}\par\vspace{5pt}}
\newcommand{\nonumsection}[1] {\vspace{12pt}\noindent{\bf #1}
	\par\vspace{5pt}}

\newcommand{\textlineskip}{\baselineskip=14pt}
\newcommand{\smalllineskip}{\baselineskip=12pt}

\def\eightcirc{
\begin{picture}(0,0)
\put(4.4,1.8){\circle{6.5}}
\end{picture}}
\def\eightcopyright{\eightcirc\kern2.7pt\hbox{\eightrm c}}

\def\abstracts#1#2#3{{
	\centering{\begin{minipage}{5in}\baselineskip=12pt\tenrm
	\centerline{ABSTRACT}
	\parindent=0pt #1\par
	\parindent=15pt #2\par
	\parindent=15pt #3
	\end{minipage} }\par}}

	{
	 \begin{list}{\arabic{enumi}.}			
	{\usecounter{enumi}\setlength{\parsep}{0pt}
	 \setlength{\leftmargin 17pt}{\rightmargin 0pt}	
	 \setlength{\itemsep}{0pt} \settowidth		
	{\labelwidth}{#1.}\sloppy}}{\end{list}}	

\newcounter{itemlistc}
\newcounter{romanlistc}
\newcounter{alphlistc}
\newcounter{arabiclistc}

\newcommand{\fcaption}[1]{
        \addtocounter{figure}{1}
         {{\tenrm Fig.~\thefigure . #1} }\hfil\break }

\newcommand{\tcaption}[1]{			
        \addtocounter{table}{1}
         {{\tenrm\offinterlineskip Table~\thetable . #1} }\hfil\break }

\def\pmb#1{\setbox0=\hbox{#1}
	\kern-.025em\copy0\kern-\wd0
	\kern.05em\copy0\kern-\wd0
	\kern-.025em\raise.0433em\box0}

\def\fnt#1#2{\footnotetext{\kern-.3em
	{$^{\mbox{\scriptsize #1}}$}{#2}}}

\def\fpage#1{\begingroup
\voffset=.3in
\thispagestyle{empty}\begin{table}[b]\centerline{\footnotesize #1}
	\end{table}\endgroup}

 1
 1
 1

\font\eightrm=cmr8

\def\qed{\hbox{${\vcenter{\vbox{                          
   \hrule height 0.4pt\hbox{\vrule width 0.4pt height 6pt
   \kern5pt\vrule width 0.4pt}\hrule height 0.4pt}}}$}}

\newcommand{\be}{\begin{eqnarray}}
\newcommand{\ee}{\end{eqnarray}}
\newcommand{\dslash}{\partial \hskip -0.5em /}
\newcommand{\Dslash}{D \hskip -0.7em /}
\newcommand{\Vslash}{V \hskip -0.7em /}

\newcommand{\Aslash}{A \hskip -0.7em /}
\newcommand{\tr}{{\rm tr}}
\newcommand{\Tr}{{\rm Tr}}

\newcommand{\La}{{\cal L}}
\newcommand{\A}{{\cal A}}

\newcommand{\ie}{{\it i.e.}\ }
\newcommand{\eg}{{\it e.g.}\ }

\input psfig

\begin{document}
\normalsize\textlineskip
{\thispagestyle{empty}
\setcounter{page}{1}

\renewcommand{\thefootnote}{\fnsymbol{footnote}} 
\def\bsc{{\sc a\kern-6.4pt\sc a\kern-6.4pt\sc a}}
\def\bflatex{\bf L\kern-.30em\raise.3ex\hbox{\bsc}\kern-.14em
T\kern-.1667em\lower.7ex\hbox{E}\kern-.125em X}

\fpage{1}

\centerline{\bf The Chiral Soliton of the Nambu--Jona-Lasinio
Model\footnote{Invited talk presented at the international workshop
on ``Baryons as Skyrme Solitons", Siegen, Sept. 92.}}
\vspace{0.2truein}
\centerline{\footnotesize REINHARD ALKOFER}
\vspace*{0.015truein}
\centerline{\footnotesize\it
Institute for Theoretical Physics, T\"ubingen University,}
\baselineskip=12pt
\centerline{\footnotesize\it
Auf der Morgenstelle 14, D-W7400 T\"ubingen, Germany }
\vspace*{0.21truein}
\abstracts{
The baryon number one soliton of the NJL model is investigated.
The Euclidian effective action is regularized using Schwinger's proper time
method. Numerical results for the static soliton
energy are presented for the cases without and with vector and
axialvector fields. It is found that the baryon number is carried by the
polarized Dirac sea of the quarks if the $a_1$ meson is included in
contrast to the case without $a_1$ where the baryon number resides on
valence quarks. Thus the model strongly supports the picture
of the baryon as a topological soliton of the meson fields.
Including the $\omega $ vector meson leads to a non--vanishing imaginary
part of the Euclidian effective action even for static solitons.
Leaving the chiral circle, i.e.\ including the isoscalar scalar meson,
the soliton first seems unstable. However, we
demonstrate that a $\phi ^4$ term stabilizes the soliton.
Finally, the self--consistent two--flavor chiral
soliton is embedded in the SU(3) flavor group and collectively quantized.
The predicted mass splittings for the low--lying baryons
with different hypercharge are
in excellent agreement with the experimental data
if the physical value for the kaon decay constant is substituted.
}{}{}

\vspace*{-3pt}
\textlineskip

\section{Introduction}

As we have learned during this workshop
there were many attempts to ``derive" the effective meson Lagrangian
from QCD in some low--energy approximation from the dynamics of the
quarks.$^1$
Unfortunately this effective meson theory is not strictly known.
However, phenomenological observations tell us that this effective
theory should embody the very successful concept of spontaneously broken
chiral symmetry which is manifest in the low energy hadron spectra.
In fact, at low energies the effective meson
Lagrangian is almost entirely determined by chiral symmetry.
Any chirally invariant quark Lagrangian, which breaks chiral symmetry
spontaneously in the vacuum, leads essentially to the same effective
meson theory. This suggests to replace (for the purpose of obtaining
the effective meson theory) QCD by a simpler chiral invariant model of
chiral flavor dynamics. In this respect the Nambu--Jona-Lasinio (NJL)
model$^2$ has been very successful. Bosonization
of this model$^3$ leads to an effective meson Lagrangian which is in
satisfactory overall agreement with low--energy meson data.
Furthermore, there are investigations which show that the NJL
model can be understood as low--energy approximation to QCD.$^4$
Given its success in the meson sector$^{5,6,7}$ it is natural to
also investigate its soliton solutions in order to describe baryons.
Unfortunately, the gradient expansion up to forth order in the
derivatives of the quark determinant fails to provide stable
soliton solutions.$^{8,9}$ Although this expansion
yields in the next to leading order the desired Skyrme term two more
terms appear which undo the job of the Skyrme term and destabilize the
chiral soliton. On the other hand, once the gradient expansion is
abandoned and the fermion determinant is treated explicitly, stable
soliton solutions are obtained.

Taking into account pseudoscalar fields only
self--consistent soliton solutions of the NJL model with full inclusion
of the Dirac sea have been found.$^{10,11,12}$ Recently, these
calculations have been extended to include vector and axialvector
mesons.$^{13,14,15,16}$ Especially, the axialvector meson $a_1$ proves to
be important. Its incorporation leads to the ``diving" of the valence
quark into the Dirac sea, and the baryons look indeed like Skyrmions.
As we will see below the scalar meson field
plays a very special role concerning the stability of the NJL
soliton.$^{17,18,19}$ In fact, the NJL model has to be suitably
generalized in order to provide stable solitons.$^{18,19}$

As is already known from the Skyrmion the static soliton does not
display the correct quantum numbers. The usual cranking procedure
projects on the nucleon and the $\Delta $.$^{20,21,22}$
Using the Yabu and Ando (YA) approach$^{23}$ to treat symmetry breaking
strange baryons as chiral
solitons of the NJL model have been studied recently.$^{24,25,26}$

This talk is organized as follows: First, the static energy functional for the
baryon number one soliton of the NJL model is discussed.
Then the question whether the scalar meson field renders the soliton unstable
is considered. Furthermore,
numerical results for the soliton of the $\pi -\rho -a_1$ and the
$\pi - \omega $ systems are presented and discussed. In the next section
the semiclassical quantization within SU(3) (and pseudoscalar fields only)
is shown. Finally, the main conclusions are summarized.

\section{The Static Energy Functional}

In this section the basic features of the underlying NJL model are
discussed. Special emphasis is hereby put on the question of
regularization  which proves to be far from trivial as the action in
terms of meson fields is in general complex, \ie it contains also an
imaginary part. The main result of this section will be the energy
functional (35) for static meson field configurations. Extremizing this
functional under the constraint of having baryon number one will yield
the static soliton.

\subsection{The Action}

The starting point for the following considerations is
the chirally invariant NJL model$^{2,3,5,6,7}$
\be
\La = \bar q (i\dslash - m^0 ) q  & + & 2g_1 \sum _{i=0}^{N_f^2-1}
\left( (\bar q \frac {\lambda ^i}{2} q )^2
      +(\bar q \frac {\lambda ^i}{2} i\gamma _5 q )^2 \right)
\nonumber \\*
 & - & 2g_2 \sum _{i=0}^{N_f^2-1}
\left( (\bar q \frac {\lambda ^i}{2} \gamma _\mu q )^2
      +(\bar q \frac {\lambda ^i}{2} i\gamma _5 \gamma _\mu q )^2 \right) ,
\ee
wherein $q$ denotes the quark spinors and $m^0$ the current quark mass
matrix.  Here we will work in the isospin limit, \ie $m_u^0=m_d^0=m^0$.
The matrices $\lambda ^i/2$ are the generators of the flavor group
($\lambda ^0 = \sqrt{2/N_f} {\bf 1}$). For the case of the
static soliton we will restrict ourselves to two flavors using $\lambda
^i =\tau ^i, \, i=0,\ldots 3.$ The coupling
constants $g_1$ and $g_2$ will be determined from mesonic properties.

Applying standard functional integral bosonization techniques the model (1)
can be rewritten in terms of composite meson fields$^3$
\be
\A &=& \A _F + \A_m,
\nonumber \\*
\A _F &=& \Tr \log (i\Dslash )
 = \Tr \log \big(i(\dslash +\Vslash +\gamma_5\Aslash )
- (P_RM+P_LM^{\dag}) \big),
\nonumber \\*
\A _m &=& \int d^4x \left( -\frac 1 {4g_1} \tr ( M^{\dag} M - m_0(M+M^{\dag}
) +m_0^2)  - \frac{1}{4g_2}\tr(V_\mu V^\mu+A_\mu A^\mu) \right) .
\nonumber \\*
\ee
Here $P_{R,L} = (1\pm \gamma _5)/2$ are the projectors on right-- and
left--handed quark fields, respectively. $V_\mu=-i
\sum_{a=0}^3V_\mu^a\tau^a/2$ and $A_\mu=-i
\sum_{a=0}^3A_\mu^a\tau^a/2$ denote the vector and axial vector meson
fields. The complex field $M$ describes the scalar and pseudoscalar
meson fields,
$S_{ij}= S^a\tau^a_{ij}/2$ and $P_{ij}= P^a\tau^a_{ij} /2$:
\be
M=S + i P = \xi_L^{\dag}\ \Phi\ \xi_R,
\ee
wherein we already introduced the angular decomposition of the complex
field $M$ into a Hermitean field
$\Phi$ and unitary fields $\xi_L$ and $\xi_R$ which are related to
the chiral field by $U=\xi_L^{\dag}\xi_R$. The latter is conveniently
expressed in terms of a chiral angle $\Theta $
\be
U(x) = \exp \left( i \Theta (x) \right) .
\ee

The quark determinant $\A_F$ diverges and therefore needs to be regularized.
As in a study of the mesonic
sector of the NJL model $^3$ as well as in previous studies of the
soliton sector $^{10,11,12}$ we will use Schwinger's proper time
regularization $^{27}$ which introduces an $O(4)$-invariant cut-off
$\Lambda$ after continuation to Euclidean space.
For the regularization procedure it is necessary to consider the real
and imaginary part of $\A_F$ independently
\be
\A_F &=& \A_R+\A_I
\nonumber \\*
\A_R &=& \frac 1 2 \Tr \log ( \Dslash_E ^{\dag}\Dslash_E )
\nonumber \\*
\A_I &=& \frac 1 2 \Tr \log ( (\Dslash_E ^{\dag})^{-1} \Dslash_E ) .
\ee
The real part $\A_R$ diverges like $\log p^2$
for large momenta $p$ whereas the imaginary part $\A_I$ does not contain
any divergencies, \ie it is finite without regularization. However, we
believe that it has to be regularized also in order to have a consistent model.
After all, the occurrence of the cutoff is a very crude way of mimicing the
asymptotic freedom of QCD.

For the real part the proper time  regularization consists in replacing
the logarithm by a parameter integral
\be
\A_R = - \frac 1 2 \int_{1/\Lambda ^2}^\infty \frac {ds}s
\Tr \exp \left( -s \Dslash_E ^{\dag}\Dslash_E \right) ,
\ee
which for $\Lambda \to \infty $ reproduces the logarithm
up to an irrelevant constant. Since the operator $\Dslash_E ^{\dag}
\Dslash_E
$ is Hermitean and positive definite this integral is well defined. For
the imaginary part the regularization procedure is equivalent,
\be
\A_I = - \frac 1 2 \int _{1/\Lambda ^2}^\infty \frac {ds}s \Tr \exp
\left( -s
(\Dslash_E ^{\dag})^{-1} \Dslash_E \right) ,
\ee
however, in this case one has to be careful concerning the convergence
of the integral. We will see later that for the cases of interest there
is no problem.

\subsection{Determination of Model Parameters}

In order to determine the coupling constants $g_1$ and $g_2$ from the
meson sector it is sufficient to only inspect $\A_R$.
Varying the regularized effective action with respect to the scalar and
pseudoscalar fields yields the Dyson--Schwinger or gap equations
\be
\langle S _{ij} \rangle &=& \delta _{ij} M_i
\nonumber \\*
M_i &=& m^0_i- 2g_1 \langle \bar q q \rangle _i
\nonumber \\*
\langle \bar q q \rangle _i &=& - M_i^3 \frac {N_c}{4\pi ^2} \Gamma
(-1,M_i^2/\Lambda ^2) .
\ee
The quantities $M_i$, $i=u,d,s$ are the dynamically generated
constituent quark masses and the $\langle \bar q q \rangle _i$ the quark
condensates  reflecting spontaneous breaking of chiral symmetry.

Employing the unitary gauge $\xi_L^{\dag}=\xi_R={\rm exp}(i\varphi)$ and
expanding the effective action up to second order in the fluctuations
$\varphi=\varphi^a(\lambda^a/2)$ allows to extract the inverse
propagator for the pseudoscalar mesons.$^{26}$
The Bethe--Salpeter equation which determines the physical meson masses
$m_\varphi$ is equivalent to the condition that the meson
propagator has a pole:
\be
D^{-1} (q^2=m_\varphi^2) =0.
\ee
The residue allows also to extract the corresponding meson decay constant
\be
f^2_{ij}(q^2)=\frac{1}{4}(M_i+M_j)^2\frac{N_c}{4\pi^2}\int_0^1dx\
\Gamma\Big(0,[(1-x)M_i^2+xM_j^2-x(1-x)m_\varphi^2]/\Lambda^2\Big).
\ee
We want to emphasize here that our Bethe--Salpeter equation is
the one in ladder approximation and that no other approximations as \eg
gradient or heat kernel expansions have been made. This also implies
that the decay constants are evaluated on the corresponding meson mass
shell. In the isospin limit ($m^0_u=m^0_d$ and therefore $M_u=M_d$)
the pion decay constant is then given
by
\be
f_\pi ^2 = M_u^2 \frac {N_c}{4\pi ^2} \int _0^1 dx
\Gamma\Big(0,[M_u^2-x(1-x)m_\pi^2]/\Lambda ^2\Big) .
\ee
One clearly sees that in the chiral limit ($m_\pi=0$) the expression
\be
f_\pi ^2 = M_u^2 \frac {N_c}{4\pi ^2}\Gamma(0,(M_u/\Lambda)^2)
\ee
calculated by means of a derivative expansion becomes exact.

For the case without vector mesons we determine the
parameters of the model, the coupling constant $g_1$, the cutoff
$\Lambda $ and the two current
masses $m_u^0$ and $m_s^0$ by fitting pseudoscalar
meson properties. First, we want
to reproduce the experimental value of the pion decay constant $f_\pi =
93$ MeV. This fixes the ratio $\Lambda /M_u$. The current masses are
determined by the pion and kaon masses, $m_\pi=135$ MeV and $m_K=495$
MeV.
This leaves one adjustable parameter, \eg the coupling constant $g_1$.
However, as $g_1$ is not very transparent we will use the gap equations
to reexpress it in terms of the up constituent mass $M_u$
which we will then choose in the range between 350 and 500 MeV. In
table~1 the corresponding values are displayed. Note that the current
masses are somewhat larger than in other regularization schemes. The
ratio $m_s^0/m_u^0$ is, as usually, varying from 23 to 25.
Note also that the kaon decay constant $f_K$ comes out about 15 \%
smaller than its experimental value $f_K = 1.24 f_\pi \approx 114$ MeV.

\begin{table}[htbp]
\tcaption{
The up and strange constituent and current masses, the cutoff
and the kaon decay constant for the parameters used later. All numbers
are in MeV.}
\centerline{\tenrm\smalllineskip
\begin{tabular}{ c c c c c c }
  \\
$M_u$  & $M_s$ & $m^0_u$  & $m^0_s$ & $\Lambda$ & $f_K$ \\
\hline
350&590&14.7&366&641&104.1\\
400&626&16.2&391&631& 99.9\\
450&663&17.1&409&633& 97.0\\
500&700&17.6&421&642& 94.9\\
\hline
\end{tabular}}
\end{table}

The inclusion of vector and axialvector mesons leads to
pseudoscalar--axialvector meson mixing, especially $\pi-a_1-$mixing.
This changes the expression for the pion decay constant. In the chiral
limit ($m^0_u=m^0_d=0$) it is given by
\be
f_\pi^2  =  \frac {6M_u^2}{g_V^2} \frac 1{1+6M_u^2/m_\rho ^2}
\ee
where
\be
m_\rho ^2 = \frac {g_V^2}{4g_2}
\quad {\rm and} \quad
g_V = \left( \frac 1 {8\pi ^2} \Gamma (0,\frac {M_u^2}{\Lambda ^2})
\right) ^{-1/2}
\ee
is the universal vector coupling constant and $M_u=M_d$ was used.
Using this procedure the cutoff $\Lambda $ is now much larger than in
the case without $\pi - a_1$--mixing, \eg for $M_u=350$MeV we have then
$\Lambda =1274$MeV instead of $\Lambda = 649$MeV if we fix $\Lambda $
from the pion decay constant $f_\pi$. $g_2$ is determined from the $\rho
$ meson mass $m_\rho$ leaving $g_1$ (or, equivalently, the constituent
mass $M_u$) as a free parameter.

\subsection{The Energy Functional for Static Meson Fields}

Next we will consider the energy functional of the static soliton.
After Wick rotation the Euclidian Dirac operator corresponding
to Eq.\ 2 is given by
\be
i\beta \Dslash_E &=& - \partial_\tau-h,
\nonumber \\*
h &=& \mbox {\boldmath $\alpha \cdot p $}+iV_4+i\gamma _5A_4 +
\mbox {\boldmath $\alpha \cdot V $} + \gamma _5 \mbox {\boldmath $\alpha
\cdot A $} +\beta (P_RM+P_LM^{\dag})
\ee
wherein $\tau$ denotes the Euclidian time. In Euclidian space
$\tau$, $V_4$ and $A_4$ have to be considered as real quantities.
This leads to a non--Hermitean Hamiltonian $h$ even for static
configurations (\ie $[\partial_\tau,h]=0$) if non--vanishing time
components of vector or axialvector meson fields are included.

For the chiral field we impose the
hedgehog ansatz:
\be
U(\mbox {\boldmath $x $})={\rm exp}\big(i{\mbox{\boldmath $\tau$}}
\cdot{\bf \hat r}\Theta(r)\Big).
\ee
This configuration has vanishing `grand spin' $\mbox{\boldmath $G$}=
\mbox{\boldmath $l$}+\mbox{\boldmath $\sigma$}/2
+\mbox{\boldmath $\tau$}/2$, \ie $[\mbox{\boldmath $G$},U]=0$.
For the scalar field only the isoscalar part may deviate from its vacuum
expectation value $M_u$. Requiring vanishing grand spin it
depends only on the radius $r$,
\be
\Phi (\mbox{\boldmath $x$}) = \phi (r) {\bf 1}.
\ee
The only possible ansatz for the $\omega$ field with grand spin zero
has vanishing spatial components($\omega_i=0$):
\be
\omega_\mu=\omega(r)\delta_{\mu4}.
\ee
For the $\rho $ and $a_1$ meson we use the spherically symmetric {\it
ans\"atze}
\be
V_\mu &=& -i \rho _\mu ^a \tau ^a, \quad \rho _0 ^a=0, \quad \rho _i^a
= -\epsilon ^{aik} \hat r^k G(r),
\nonumber \\
A_\mu &=& -i a_\mu ^a \tau ^a, \quad a_0^a=0, \quad
a_i^a = \hat r^i \hat r^a F(r) + \delta ^{ia} H(r)
\ee
where the index $a$ runs from 1 to 3.

Thus the Euclidian Dirac Hamiltonian reads
\be
h &=& \mbox {\boldmath $\alpha \cdot p $}+i\omega(r)
+ \phi (r) \beta({\rm cos}\Theta(r)+i\gamma_5
{\mbox{\boldmath $\tau$}}\cdot{\bf \hat r}{\rm sin}\Theta)
\nonumber \\*
&+&\frac 1 2 (\mbox {\boldmath $\alpha $} \times \hat r ) {\mbox{\boldmath
$\tau$}} G(r)
+ \frac 1 2 (\mbox {\boldmath $\sigma \cdot $} \hat r )
(\mbox {\boldmath $\tau  \cdot $} \hat r ) F(r)
+\frac 1 2 (\mbox {\boldmath $\sigma \cdot \tau $} ) H(r).
\ee
which obviously is not Hermitean since $\omega(r)$ is real giving
rise to complex eigenvalues of $h$. For static
configurations the eigenvalues of $\partial_\tau$,
$i\Omega_n=i(2n+1)\pi/T, \  (n=0,\pm1,\pm2,..)$  may be
separated.\footnote{The eigenfunctions of $\partial_\tau$ assume anti-periodic
boundary conditions in the Euclidian time interval $T$. The
$\Omega_n$ are the analogues of the Matsubara frequencies with $T$
figuring as inverse temperature.}
{}~Thus the eigenvalues
$\lambda_{n,\nu}$ of the operator $\partial_\tau+h$ read:
\be
\lambda_{n,\nu}=-i\Omega_n+\epsilon_\nu=-i\Omega_n+
\epsilon_\nu^R+i\epsilon_\nu^I.
\ee
The fermion determinant is expressed in terms of the eigenvalues
$\lambda_{n,\nu}$:
\be
\A_R=\frac{1}{2}\sum_{\nu,n}{\rm log}(\lambda_{n,\nu}\lambda_{n,\nu}^*)
\qquad {\rm and}\qquad
\A_I=\frac{1}{2}\sum_{\nu,n}{\rm log}(\frac{\lambda_{n,\nu}}
{\lambda_{n,\nu}^*}).
\ee
Using (21) the real part reads:
\be
\A_R&=&\frac{1}{2}\sum_{\nu,n}{\rm log}\big((\Omega_n-\epsilon_\nu^I)^2+
(\epsilon_\nu^R)^2\big) \nonumber \\
&&\qquad\to\ -\frac{1}{2}\sum_{\nu,n}\int_1^\infty
\frac{d\tau}{\tau} {\rm \exp}\big\{-\frac{\tau}{\Lambda^2}
\big((\Omega_n-\epsilon_\nu^I)^2+(\epsilon_\nu^R)^2\big)\big\}
\ee
according to the proper time regularization scheme (6). For large
Euclidian time intervals ($T\rightarrow\infty$) the temporal part of
the trace may be performed
\be
\A_R=-\frac{T}{2}\sum_\nu\int_{-\infty}^{\infty}\frac{dz}{2\pi}
\int_1^\infty \frac{d\tau}{\tau}
{\rm \exp}\big\{-\frac{\tau}{\Lambda^2}\big(z^2
+(\epsilon_\nu^R)^2\big)\big\}
\ee
where we have shifted the integration variable $z-\epsilon_\nu^I
\rightarrow z$. For $T\rightarrow\infty$ we may read off the
Dirac sea contribution to the real part of the energy functional
from $\A_R\rightarrow -TE_{\rm vac}^R$:
\be
E_{\rm vac}^R=\frac{N_C}{4\sqrt{\pi}}\sum_\nu |\epsilon_\nu^R|
\Gamma\big(-\frac{1}{2},(\epsilon_\nu^R/\Lambda)^2\big).
\ee
For soliton configurations with vanishing $\omega$ (\ie
$\epsilon_\nu^R=\epsilon_\nu$) there is no contribution from the
imaginary part and eq.\ (25) is the expression
for the energy of the Dirac sea.

For the imaginary part we obtain
\be
\A_I=\frac{1}{2}\big(\sum_\nu\sum_{n=-\infty}^\infty
{\rm log}(\lambda_{\nu,n})-
\sum_\nu\sum_{n=-\infty}^\infty{\rm log}(\lambda_{\nu,n}^*)\big)=
\frac{1}{2}\sum_\nu\sum_{n=-\infty}^\infty{\rm log}
\frac{i\Omega_n-\epsilon_\nu}{i\Omega_n-\epsilon_\nu^*}
\ee
where we have reversed the sign in the first sum over the integer
variable $n$. Next we express $\A_I$ in terms of a parameter integral:
\be
\A_I=\frac{1}{2}\sum_\nu\sum_{n=-\infty}^\infty
\int_{-1}^1 d\lambda \frac{-i\epsilon_\nu^I}
{i\Omega_n-\epsilon_\nu^R-i\lambda\epsilon_\nu^I}.
\ee
In analogy to (24) we may carry out the temporal trace in the
limit $T\rightarrow\infty$:
\be
\A_I=\frac{-i}{2}\sum_\nu \int_{-1}^1 d\lambda\, T
\int_{-\infty}^{\infty}\frac{dz}{2\pi}\epsilon_\nu^I
\big[i(z-\lambda\epsilon_\nu^I)-\epsilon_\nu^R\big]^{-1} .
\ee
Shifting the integration variable $z-\lambda\epsilon_\nu^I
\rightarrow z$ the integral over $\lambda$ may be done
\be
\A_I=\frac{-i}{2}\sum_\nu \epsilon_\nu^I
\int_{-\infty}^{\infty}\frac{dz}{2\pi}
\frac{-2\epsilon_\nu^R}{z^2+(\epsilon_\nu^R)^2}.
\ee
$\A_I$ is regularized in proper time by expressing the integrand as
a parameter integral:
\be
\frac{-1}{z^2+(\epsilon_\nu^R)^2}\to\int_{1/\Lambda^2}^\infty d\tau
{\rm exp}\big\{-\tau(z^2+(\epsilon_\nu^R)^2\big\}
\ee
which obviously is finite for $\Lambda\rightarrow\infty$. Continuing
the evaluation of $\A_I$ in analogy to eqs.\ (23-25) we find for the
contribution of the Dirac sea to the imaginary part of the
Euclidian energy
$E_{\rm vac}^I$
\be
E_{\rm vac}^I=\frac{-N_C}{2}\sum_\nu \epsilon_\nu^I {\rm sign}
(\epsilon_\nu^R)\cases{1,&$\A_I\quad {\rm not}\quad {\rm
regularized}$\cr
{\cal N}_\nu, &$\A_I\quad {\rm regularized}$\cr}
\ee
where ${\cal N}_\nu = \frac{1}{\sqrt\pi}\Gamma\big(\frac{1}{2},
(\epsilon_\nu^R/\Lambda)^2\big)$ are the vacuum ``occupation
numbers" in the proper time regularization scheme.
The upper case, of course, corresponds to the limit
$\Lambda\rightarrow\infty$. Obviously only the real part of the
one-particle energy eigenvalue is relevant for the regularization
of $\A_I$. Eq.\ (31) reveals that we have succeeded in finding a
regularization scheme for $\A_I$ that only involves quantities which
are strictly positive definite. This is not evident from
the definition of $\A_I$ (5).

The total energy functional contains besides $E_{\rm vac}^R$ and
$E_{\rm vac}^I$ also the valence quark energy
\be
E_{\rm val}^R=N_C\sum_\nu\eta_\nu |\epsilon_\nu^R|\qquad
E_{\rm val}^I=N_C\sum_\nu\eta_\nu {\rm sign}(\epsilon_\nu^R)
\epsilon_\nu^I
\ee
with $\eta_\mu=0,1$ being the occupation numbers of the valence quark
and
anti-quark states.
Furthermore the meson energy is obtained by substituting the
ans\"atze (16-19) into (2):
\be
E_m&=&4\pi \int dr r^2\Bigl( \frac {\langle \bar uu \rangle }{M_u-m^0_u}
(\phi ^2(r)-M_u^2) + m_\pi^2f_\pi^2(1-\frac {\phi (r)}{M_u}
\cos\Theta(r))
\nonumber \\*
&& \qquad
+af_\pi^2 \bigl( G^2(r)+\frac 1 2 F^2(r) + F(r) H(r) + \frac 3 2 H^2 (r)
-2 \omega^2(r) \bigr) \Bigr)
\ee
where $a=1+m_\rho^2/6M_u^2$ if $\pi -a_1-$mixing is included and
$a=m_\rho^2/6M_u^2$ if it is not. Note that we are working in the
isospin limit which implies $m_\omega = m_\rho$.
Continuing back to Minkowski space we find for the total energy
functional:
\be
E[\phi ,\Theta,\omega , G,F,H]
=E_{\rm val}^R+E_{\rm val}^I+E_{\rm vac}^R+E_{\rm vac}^I+E_m.
\ee

The equations of motion are now obtained by extremizing the static
Minkowski energy. They are  solved by iteration. We start off with
test meson profiles to calculate the matrix elements of the Hamiltonian.
The resulting eigenvalues
and eigenvectors are substituted into the equations of motion
yielding new profile functions. This process is repeated until
convergence is reached.
If $h$ is not Hermitean (in Euclidian space) which is
the case if the $\omega$ meson is included one has
to distinguish between left and right eigenvectors of $h$. For more
details see ref.\ $^{16}$. Otherwise $h$ can be written in a real and
symmetric form using eigenfunctions of the Hamiltonian with vanishing
meson fields and of the grand spin as basis. The momenta are
discretized using a finite spherical box. An important point are the
boundary conditions used. A discussion of the advantages of two types of
boundary conditions may be found in ref.\ $^{26}$.

\section{Instability due to scalar mesons?}

The first calculations have been done with restriction to the chiral
field only $^{10,11,12}$. Then one finds stable solitons for constituent
masses $M_u> 325$MeV where $f_\pi =93$MeV is used to fix the cutoff.
The static soliton mass is approximately 1200 MeV almost independent of
$M_u$. For physical values of $M_u$, $M_u\approx 350$MeV, the valence quark
energy is positive and half of the soliton energy is carried by the
valence quark. Even more, quantities like the moments of inertia are
dominated by the valence quark contribution.$^{26}$

Restricting oneself to the pseudoscalar field only the stability of the
soliton can be easily understood. The value of the chiral angle at the
origin is fixed, $\Theta (r=0) = \pi $. Assuming $\Theta $ to
be monotonically decreasing it can be characterized by its spatial
extension. The valence quark favors a large
spatial extension of the soliton because it then becomes stronger bound,
\ie its energy is decreased. On the other hand, the Dirac sea is much
less distorted for small soliton sizes and therefore its contribution to
the energy behaves oppositely to the one of the valence quark. Given
both features one obtains a minimum in the energy.

Recently it has been shown that including the scalar isoscalar
field renders the soliton unstable.$^{17}$ Allowing this field to
develop a narrow and high peak near the origin leads to the collapse of
the NJL soliton. Note that this kind of instability is qualitatively
different to the ``breathing mode instability"
which could have occurred for the case of the chiral
field only (and which is the one discussed for Skyrme type models).
Especially, the narrow and high peak which develops in the scalar field
leads to the fact that very high momentum components of this field are
involved.

For a number of reasons, the scalar--isoscalar sector of the NJL model is the
least well understood of the effective dynamics. The absence of clearly
interpretable
resonances in this channel makes it difficult to determine parameters
of effective models in the meson sector.\footnote{Additionally, the
physical scalar mesons mix with states composed of two pseudoscalar
mesons making the extraction of the properties of ``bare" scalars
extremely complicated.}
Moreover, it is likely that $\bar qq$--scalars strongly
mix with scalar glueballs, whose presence in the low--energy dynamics is
required by the scale anomaly of QCD. In an effective Lagrangian
language these are described by a scalar field which is subject to an
effective potential leading to a non--vanishing vacuum value. The feature
which interests us here most is the fact that the mixing of the scalar
glueball with the scalar meson $\Phi $ leads to a $\Phi ^4$ term in the
action. As will be demonstrated in the following such a term prevents
the collapse of the soliton.$^{18,19}$

Employing the special parametrization $\Phi U=\sigma+
{\mbox{\boldmath $\tau$}}\cdot{\mbox{\boldmath $\pi$}}$ we use the {\it
ansatz}
\be
\sigma(r)=M_u\big[1+Wf(r/R){\rm cos}\Theta(r/R)\big],\quad
{\mbox{\boldmath $\pi$}}({\bf r})={\bf \hat r}
M_u W f(r/R){\rm sin}\Theta(r/R)
\ee
with $W(RM_u)^\alpha=const$. $R$ denotes a characteristic radius
which parametrizes the size of the field configuration. In ref.$^{17}$
a Wood-Saxon type shape was chosen for $f(r/R)$ while
a straight line profile was assumed for $\Theta(r/R)$.
The constant $\alpha$ is determined such that the mesonic
part of the energy vanishes in the limit $R\rightarrow0$
\be
\alpha<\cases{\frac{3}{2} & {\rm for\ a\ $\phi^2\ $ potential}\cr
\frac{3}{4} &  {\rm for\ a\ $\phi^4\ $ potential}}
\ee

In figure 1 we display the contribution of the fermion determinant
to the static energy. Obviously the meson profile collapses in
case there is only a $\phi^2$ potential, however, in the presence
of a  $\phi^4$ term in the mesonic energy functional the fermion
determinant is strongly repulsive and prevents the soliton from
collapsing.

\begin{figure}[t]
\centerline{\ \psfig{figure=scal1.ps,height=10cm,width=16cm}}
\fcaption{The energy of the fermion determinant for two
different meson field configurations. Note that the mesonic part
of the energy vanishes for $R\rightarrow0$ in both cases.}
\end{figure}

\begin{figure}
\centerline{\ \psfig{figure=scal2.ps,height=9cm,width=12cm}}
\centerline{\ \psfig{figure=scal3.ps,height=9cm,width=12cm}}
\fcaption{The scalar meson profile $\phi (r)$ and the chiral angle
$\theta (r)$ for the non--linear case (full line), for $a/b=1$ (dashed
line) and $a/b=1/10$ (dotted line).}
\end{figure}

To be more explicit we calculated the self--consistent soliton including
a contribution to the mesonic energy of the form
\be
a\phi^4(r) + bM_u^2 \phi^2(r)
\ee
where the ratio between
the parameters $a$ and $b$ (the unconventional coefficient $bM_u^2$
was chosen in order to have dimensionless parameters) has been determined
by fixing simultaneously the constituent mass and the quark condensate.
The cutoff $\Lambda $ was, as usual, fixed using the pion decay constant
as input. The profile function $\phi (r)$ as well as the chiral angle
$\Theta $ are compared for two different values of $a/b$ to some older
calculation using pseudoscalar fields only are shown in figure 2. One
clearly sees that $\Theta (r)$ does not change much even if $a/b$ is as
small as 1/10. Note that $a\to \infty$ corresponds to the non--linear
case, \ie to the restriction to the chiral circle. The soliton energy
decreases from 1214 MeV in the non--linear case to 1179 MeV for the case
$a/b=1/10$, \ie we have only a three percent decrease even for very
small $a/b$. Therefore one may conclude that the NJL soliton based on a
generalized model with $\Phi ^4$ term is very similar to the one where
the scalar field is constrained to its vacuum value.

\section{The soliton of the $\pi - \rho - a_1 -$system}

In the following
we will use a fixed $\phi (r)=M_u$. However, from eq.\ (3) it is
obvious that also vector and axialvector meson fields are present in the
soliton for which we will use the profile functions of eqs.\ (18) and
(19). Postponing the difficulties which arise due to the imaginary part
of the action generated by the $\omega $ meson to the next section we
restrict ourselves for the moment to the $\pi - \rho - a_1 -$system.

Additionally, we consider only the chiral limit $m_u^0=m_d^0=0$.
The cutoff $\Lambda $ and the coupling constant $g_2$ are determined as
described in section 2.2.
As input we use $f_\pi = 93$MeV and $m_\rho =770$MeV, see eqs.\ (13)
and (14). In the numerical method we used two different boundary
conditions (for a detailed description see appendix B of ref.$^{26}$)
in order to be sure that spurious contributions are negligible.
For the cases $M_u=300$, 350 and 400 MeV the total soliton energy
$E$ as well as its different
contributions are shown in table~2. The energy eigenvalue of the
valence quark is negative in all three cases. Its contribution
is therefore already fully included in the vacuum
part of the soliton energy. Note that there is also no (direct)
contribution of the valence quark
to other physical variables, especially the baryon current:
{\bf The baryon number is carried by the vacuum.}
As the vacuum is unobservable this means that the baryon number
``resides" in the meson profiles, \ie these results support
Witten's conjecture that baryons may be described as solitons within
purely mesonic models.

\begin{table}[htbp]
\tcaption{
The soliton energy $E$ as well as its Dirac sea and mesonic
contributions $E_{sea}$ and $E_{mes}$ for different values of the
constituent quark mass $M$. Shown is also the energy of the 'dived'
level.}
\centerline{\tenrm\smalllineskip
\begin{tabular}{l r r r}
\\
$M$ (MeV)       & 300 & 350& 400 \\
\hline
$E$ (MeV)       &1104 &1010 & 938 \\
$E_{sea}$ (MeV) & 736 & 615 & 544 \\
$E_{mes}$ (MeV) & 368 & 395 & 394 \\
$e_{val}/M$     & -0.04 &-0.38 &-0.54\\
\hline
\end{tabular}}
\end{table}

\begin{table}[htbp]
\tcaption{The contributions to the soliton energy for the case where one
considers the chiral angle $\Theta $ only, with $\Theta $ and $\rho $
and with $\Theta $,$\rho $ and $a_1$ for a constituent mass $M=350$MeV.
For completeness the values of the cutoff $\Lambda $ are shown.}
\centerline{\tenrm\smalllineskip
\begin{tabular}{l r r r }
\\
 &only $\Theta $ & $\Theta $ and $\rho $ & $\Theta $,$\rho $ and $a_1$ \\
\hline
$E$ (MeV)       & 1214 & 957 & 1010 \\
$E_{val}$ (MeV) &  653 & 147 &    0 \\
$E_{sea}$ (MeV) &  561 & 655 &  615 \\
$E_{mes}$ (MeV) &    0 & 155 &  395 \\
$\Lambda $ (MeV) & 649 & 649 & 1274 \\
\hline
\end{tabular}}
\end{table}

\begin{figure}
\centerline{\ \psfig{figure=vecth.ps,height=5cm,width=12cm}}
\vskip 1cm
\centerline{\ \psfig{figure=vecrho.ps,height=5cm,width=12cm}}
\vskip 1cm
\centerline{\ \psfig{figure=veca1.ps,height=5cm,width=12cm}}
\vskip 1cm
\fcaption{The chiral angle $\Theta $, the $\rho $ meson profile $G(r)$
and the $a_1$ meson profiles $F(r)$ and $H(r)$ for the three cases of
table 3.}
\end{figure}

In table 3 we compare the
contributions to the soliton energy for three types of calculations:
In the first only the pseudoscalar field is taken into account, in the
second one includes the $\rho $ meson and in the third one the $\pi -
\rho -a_1$-system is considered. One sees clearly that the presence of
the $a_1$ field leads to the ``diving" of the valence quark level. This
can be easily understood from the $a_1$ meson profiles, see figure 3.
As the valence quark level is a $s$--wave with respect to grand spin it
will be mostly influenced by meson profiles which are non--vanishing at
the origin as $H(r)$ does in contrast to $G(r)$ and $F(r)$. Furthermore,
as can be seen from figure 3 the chiral angle is ``squeezed" if the
$a_1$ meson is included.

We want to emphasize here that the NJL soliton including the
$\rho$ and $a_1$--mesons in addition to the chiral field supports
Witten's conjecture that baryons may be described as solitons within
purely mesonic models. Earlier calculations without the $a_1$--meson
seemed to indicate that baryonic properties are dominated by valence
quarks in contradiction to the assumption of a purely mesonic
topological
soliton description of baryons. This situation is in agreement with
the experience one has with the pure Skyrme model where only the chiral
field is included and the (axial-) vector mesons are left out. One knows
that the Skyrmion based on pseudoscalars only has serious deficiencies,
as \eg wrong electromagnetic properties due to missing vector dominance
$^{28}$
or wrong "high energy behavior" of $\pi - N$ phase shifts due to the
higher order stabilization terms.$^{29}$

\section{The soliton of the $\pi -\omega -$system}

We have calculated the chiral soliton with the $\omega$ vector meson
for a wide range of values for the constituent mass.
In contrast to the results of ref.$^{15}$  no instability is obtained,
especially we find stable soliton solutions for the parameters
which are extracted from mesonic data (see eqs.\ 13,14). This difference
is most likely due to the fact that in ref.$^{15}$ eigenvalues of
Minkowski space Hamilton operators have been used in order to regularize.
However, regularization can be defined only in Euclidean space, see
eqs.\ 6 and 7. Therefore the treatment of ref.$^{15}$ is inconsistent. A
more detailed discussion can be found in ref.$^{16}$.

\begin{figure}
\centerline{\ \psfig{figure=ompl3.ps,height=6cm,width=12cm}}
\fcaption{The total static energy of the soliton as a function of the
constituent mass. The various contributions are: real part (dotted),
imaginary part (dashed-dotted) and the mesonic part (dashed).}
\end{figure}

\begin{figure}
\centerline{\ \psfig{figure=omyl.ps,height=6cm,width=12cm}}
\fcaption{The radial dependence of the $\omega$ meson profile for
the constituent mass $M_u=600$MeV. The imaginary part is regularized.
The dashed line denotes the real part and the dotted line the imaginary
part while the solid line represent their sum.}
\end{figure}

{}From figure 4 we see that the soliton energy (34) varies only mildly
as a function of the constituent mass. The decrease of the
real part is compensated by the increase of the imaginary part. The
mesonic part is almost independent of the constituent mass.
Noting that the energy of the NJL soliton with the chiral field only
typically is of the order $1.2$GeV $^{10,11,12}$ we recognize that the
inclusion of the $\omega$ meson increases the total energy.
Actually we find that the real part is lowered by the inclusion
of $\omega$, however, the additional imaginary part pushes the
result for the total energy above the value obtained in the
soliton model with only $\Theta$. This indicates
that equations of motion for time components of vector fields
prove to be constraints which is realized by the complex nature of the
energy in Euclidian space.

Furthermore, the results demonstrate that the non-regularization
of $E_{\rm vac}^I$ does not strongly effect the result for the total
energy.
Merely the increase of the imaginary part with $M_u$ is somewhat more
pronounced since the exponential damping in (31) drops.
However, the (non-) regularization of
$E_{\rm vac}^I$ has a significant effect on the $\omega$-meson profile;
especially we find that for $M_u \ge 600$MeV $\omega(r)$ does not
assume its maximal value at the origin when $E_{\rm vac}^I$ is not
regularized. We furthermore recognize that there exists a large
cancellation between the real part of $\omega(r)$ and its imaginary
part, see figure 5. In contrast to the na\"\i ve
expectation we find for small $M_u$ that the real part of the baryon
density dominates $\omega(r)$ in the vicinity of the origin.

Of course, the final goal is to incorporate all vector mesons in the
calculation of the soliton of the NJL model and work in this direction
is in process. This implies a fusion of the work presented in this and
the preceding section.
Actually we do not expect the conclusion that the whole baryon
charge is carried by the asymmetry of the polarized Dirac sea to change
since including the $\omega$ meson we
find that the critical constituent mass at which the real part of
the valence quarks' energy changes sign $M_{\rm crit.}=545$MeV
is already about 180MeV lower than in the NJL model including
the chiral field only.$^{10,11,12}$

\section{Semiclassical Quantization in SU(3)}

The chiral hedgehog soliton has neither the good spin nor flavor quantum
numbers of the physical baryons. To project on baryon quantum
numbers we employ the cranking procedure and impose the following
ansatz:
\be
\xi(\mbox{\boldmath $x$},t)=R(t)\xi(\mbox{\boldmath $x$})
R^{\dag}(t)
\ee
where $R(t)$ describes the (adiabatic) rotation in $SU(3)$ flavor space.
Of course, we only have zero modes in the subspace $SU(2)_I\times
U(1)_Y$ of $SU(3)$, nevertheless (38) is reasonable since we consider
$SU(3)$ as an approximate symmetry. Elevating $R(t)$ to be $SU(3)$
valued furthermore allows us to easily make contact with the
phenomenology of baryon representations. This approach has received
intensed recognition after Yabu and Ando (YA)$^{23}$ demonstrated that the
resulting collective Hamiltonian may be diagonalized exactly. More
recently the YA approach has been extented considerably to more
complicated symmetry breaking schemes.$^{30,31,32}$

The main purpose of the investigations in this section is to explain
how semiclassical quantization can be applied to the NJL soliton and
furthermore provides reasonable results for baryon mass
differences. Therefore we shall restrict ourselves to the
approach which contains pseudoscalar fields only.

\vfil\eject

\subsection{Description of the approach}

In the following it will be convenient to transform the
fermion determinant to the flavor rotation frame
$q(\mbox{\boldmath $x$},t)=R(t) q^\prime(\mbox{\boldmath $x$},t)$.
Obviously this transformation cancels the rotation matrices in our
{\it ansatz} (38) at the expense of induced terms in the quark Hamiltonian
due to the time dependence of the rotations and symmetry breaking:
\be
h=h_0+h_{rot}+h_{SB}
\ee
wherein $h_0$ is the static $SU(2)$ quark Hamiltonian introduced in
section 2. The induced Hamiltonian $h_{rot}$ originates from the time
dependence of the collective rotation
\be
h_{rot}=-iR^{\dag}(t){\dot R}(t)=
\frac{1}{2}\sum_{a=1}^8\lambda_a\Omega_a=
\pmatrix{{\Omega_\pi}+{\Omega_\eta}&\Omega_K\cr
{\Omega_K^{\dag}}&-2{\Omega_\eta}\cr}.
\ee
The contribution to $h$ due to $SU(3)$ symmetry breaking is more
difficult to handle:
\be
h_{SB}&=&{\cal T}\beta\big[R^{\dag}(t)\langle S\rangle R(t)-
\langle S\rangle\big]{\cal T}^{\dag} \nonumber \\
&=&\frac{\Delta M}{\sqrt3} {\cal T} \big[\beta
\sum_{i=1}^3\big(D_{8i}\lambda_i+
\sum_{\alpha=4}^7D_{8\alpha}\lambda_\alpha+
(D_{88}-1)\lambda_8\big)\big] {\cal T}^{\dag}
\ee
with ${\cal T}=\xi(\mbox{\boldmath $x$})P_L
+\xi^{\dag}(\mbox{\boldmath $x$})P_R$ and $\Delta M =M_u-M_s$
denotes the difference of the up and strange constituent masses.
The adjoint representation
$D_{ij}=\frac{1}{2}\tr(\lambda_iR\lambda_jR^{\dag})$ of the rotation
matrices clearly exhibits the transformation properties of the symmetry
breaking part of the Hamiltonian.

Our goal is to expand the energy in both $\Delta M$ and
the angular velocities $\Omega_a$. The latter actually corresponds to
an expansion in $1/N_C$. In this expansion we keep terms quadratic in
the perturbation $h_P=h_{rot}+h_{SB}$. Then, due to isospin invariance,
the most general form of the collective Lagrangian reads:
\be
L=&-&E+\frac{1}{2}\alpha^2\sum_{i=1}^3\Omega_i^2+
\frac{1}{2}\beta^2\sum_{\alpha=4}^7\Omega_\alpha^2-
\frac{\sqrt3}{2}B\Omega_8+\alpha_1\sum_{i=1}^3D_{8i}\Omega_i
+\beta_1\sum_{\alpha=4}^7D_{8\alpha}\Omega_\alpha\nonumber \\*
&-&\frac{1}{2}\gamma(1-D_{88})-\frac{1}{2}\gamma_1(1-D_{88}^2)
-\frac{1}{2}\gamma_2\sum_{i=1}^3D_{8i}D_{8i}
-\frac{1}{2}\gamma_3\sum_{\alpha=4}^7D_{8\alpha}D_{8\alpha}.
\ee
$E$ hereby is the classical soliton energy. In the following
we will describe the origin of all terms in the collective Lagrangian.

The valence quarks are  treated by standard perturbation techniques,
however, they are only explicitly added as long as the corresponding
energy eigenvalue is positive. Due to the embedding of the static
soliton solution in the $SU(2)$ subgroup only iso-singlet operators
contribute non-trivially in first order perturbation theory. Since
$\lambda_8$ acts like $\frac{1}{\sqrt3}\times{\rm unit-matrix}$ in the
isospin subgroup we obtain:
\be
B^{\rm val}=\eta^{\rm val}
\ee
which is the valence quark contribution to the baryon number.
By the same argument
\be
\gamma^{\rm val}=-\frac{2}{3}N_C\Delta M \eta^{\rm val}
\langle{\rm val}|{\cal T}\beta{\cal T}^{\dag}|{\rm val}\rangle.
\ee
The valence quark contribution to the expressions quadratic in angular
velocities, $\frac{1}{2}\Theta_{ab}\Omega_a \Omega_b$ is the well known
cranking result:
\be
\Theta^{\rm val}_{ab}=\frac{N_C}{2}\eta^{\rm val}\sum_{\mu\ne {\rm val}}
\frac{\langle{\rm val}|\lambda_a|\mu\rangle \langle\mu|\lambda_b|{\rm
val}\rangle} {e_\mu-e_{\rm val}}.
\ee
Isospin invariance demands that the matrix for the moment of inertia,
$\Theta^{\rm val}_{ab}$ is diagonal and only two independent components,
$\Theta^{\rm val}_{33}$ and $\Theta^{\rm val}_{44}$ corresponding to
$\alpha^2$ and $\beta^2$ respectively, exist. The same argument, of
course, applies to the terms quadratic in $\Delta M$
contributing to $\gamma_2$ and $\gamma_3$ as well:
\be
\Gamma^{\rm val}_{ab}=\frac{2}{3}N_C(\Delta M)^2
\eta^{\rm val}\sum_{\mu\ne {\rm val}}
\frac{\langle{\rm val}|{\cal T}\beta\lambda_a{\cal T}^{\dag}|\mu\rangle
\langle\mu|{\cal T}\beta\lambda_b{\cal T}^{\dag}|{\rm val}\rangle
}{e_\mu-e_{\rm
 val}}
\ee
and also to those linear in both $\Omega_a$ and $\Delta M$:
\be
\Delta^{\rm val}_{ab}=\frac{N_C}{\sqrt3}\Delta M
\eta^{\rm val}\sum_{\mu\ne {\rm val}}
\frac{\langle{\rm val}|\lambda_a|\mu\rangle \langle\mu|{\cal
T}\beta\lambda_b
{\cal T}^{\dag}|{\rm val}\rangle }{e_\mu-e_{\rm val}}  .
\ee
$\Delta^{\rm val}_{ab}$ contributes to $\alpha_1$ and $\beta_1$.
In all three cases the 88-component vanishes since the classical
hedgehog commutes with $\lambda_8$.

Next we turn to the evaluation of the vacuum contribution to the
fermion determinant in presence
of the perturbation $h_P$. As for the calculation of the static soliton
solution we transform to Euclidean space and take the vacuum
$(\xi=1)$ as reference:
\be
\A_F= \Tr \log \big(i\Dslash_E(\xi(\mbox{\boldmath $x$},t))\big)-
\Tr \log \big(i\Dslash_E(\xi=1)\big).
\ee
In the flavor rotating frame we have
\be
\bar q \Dslash_E q = \bar q^{\prime} \Dslash\ ^{\prime}_E q^{\prime},
\qquad
{\rm with}\qquad
i\Dslash\ ^{\prime}_E=\beta\big(\partial_\tau -
(h_0+h_{rot}+h_{SB})\big).
\ee
Attention has to be paid to the fact that in Euclidean space the
angular velocities $\Omega_a$ are to be considered anti-Hermitean
quantities.

Again it proves most convenient to treat real and imaginary parts
separately. First we consider the real part in the flavor rotation
frame:
\be
\A_R=-\frac{1}{2}\int_{1/\Lambda^2}^\infty \frac{ds}{s} \Tr \exp
(-s \Dslash\ ^{\prime{\dag}}_E \Dslash\ ^{\prime}_E).
\ee
This quantity is expanded up to second order in $h_P$.
Here we will confine ourselves to the presentation of the resulting
coefficients in the collective Lagrangian. Their derivation is presented
in appendix C of ref.$^{26}$.

The underlying assumption of all these calculations is that of a
strictly adiabatic rotation. Then we recover the old result
for the moments of inertia:
\be
\Theta^{\rm vac}_{ab}=\frac{N_C}{4}\sum_{\mu\nu}
f_\Theta(e_\mu,e_\nu;\Lambda)\langle\mu|\lambda_a|\nu\rangle
\langle \nu|\lambda_b|\mu\rangle
\ee
where the cut-off function $f_\Theta(e_\mu,e_\nu;\Lambda)$ is given
by
\be
f_\Theta(e_\mu,e_\nu;\Lambda)=\frac{1}{\Lambda\sqrt{\pi}}
\frac{e^{-E_\mu^2}-e^{-E_\nu^2}}{E_\nu^2-E_\mu^2}-
\frac{2}{\Lambda}\frac{1}{E_\mu-E_\nu}
\big[{\rm sign}(E_\nu){\cal N}_\mu-
{\rm sign}(E_\mu){\cal N}_\nu\big]
\ee
with $E_\mu=e_\mu/\Lambda$.
Due to different up  and strange constituent masses the strange
moment of inertia $\Theta^{\rm vac}_{44}$ does not vanish for
$\xi=1$ and the corresponding subtraction (48) has to be
performed.

The expansion in $\Delta M$ yields a linear contribution
\be
\gamma^{\rm vac}=-\frac{2N_C\Delta M}{\sqrt3}\sum_\mu {\rm sign}(e_\mu)
{\cal N}_\mu \langle \mu|{\cal T}\beta\lambda_8{\cal
T}^{\dag}|\mu\rangle
\ee
wherein after subtracting the $\xi=1$ result only the sum over
non-strange
states remains. The term quadratic in $\Delta M$ reads
\be
\Gamma^{\rm vac}_{ab}=\frac{N_C}{3}(\Delta M)^2\sum_{\mu\nu}
f_\Gamma(e_\mu,e_\nu;\Lambda)
\langle \mu|{\cal T}\beta\lambda_a{\cal T}^{\dag}|\nu\rangle
\langle \nu|{\cal T}\beta\lambda_b{\cal T}^{\dag}|\mu\rangle
\ee
with the cut-off function
\be
f_\Gamma(e_\mu,e_\nu;\Lambda)=\frac{{\rm sign}(e_\mu){\cal N}_\mu
- {\rm sign}(e_\nu){\cal N}_\nu}{e_\mu-e_\nu}.
\ee

The expansion of the real part of the action does not provide any term
linear in the angular velocities. This may be understood by noting that
an expansion of $\A_R$ only yields terms even in time derivatives.
However, these linear terms are provided by the imaginary part
since it is known from the derivative expansion of $\A_I$ that
only odd powers of the time derivative operator appear.

In contrast to the real part of the Euclidean action the imaginary
part, which in the flavor rotating frame reads:
\be
\A_I=\frac{1}{2}\Tr \log \big((\Dslash\ ^{\prime{\dag}}_E)^{-1}
\Dslash\ ^\prime_E\big)
\ee
vanishes for $\xi=1$ and therefore no vacuum subtraction is needed.
We get contributions from $\A_I$ to the baryon number
\be
B^{\rm vac}=\sum_\mu {\rm sign}(e_\mu){\cal N}_\mu
\ee
and to the symmetry breaker linear in $\Delta M$:
\be
\Delta^{\rm vac}_{ab}=\frac{-N_C}{2\sqrt3}\Delta M\sum_{\mu\nu}
f_\Gamma(e_\mu,e_\nu;\Lambda)\langle \mu|\lambda_a|\nu\rangle
\langle \nu|{\cal T}\beta\lambda_b{\cal T}^{\dag}|\mu\rangle .
\ee

Let us perform a small detour to Skyrme type, \ie purely mesonic,
models. There it is well known that kaon fields are induced by the
collective rotation into strange direction. Parametrizing$^{33}$
\be
\xi(\mbox{\boldmath $x$},t)=R(t)\xi_k\xi(\mbox{\boldmath $x$})
\xi_k R^{\dag}(t);\ \xi_k={\rm exp}(iZ)
\ee
the induced kaon fields are contained in $Z$:
\be
Z=\pmatrix{0 &K\cr K^{\dag}&0\cr}\quad,\qquad
K=W(r)\sum_{i=1}^3(\hat x_i\lambda_i)\Omega_K.
\ee
The strange moment of inertia, $\beta^2$, turns out to be a functional
of the radial function $W(r)$. The inclusion of the Wess-Zumino term
$\Gamma_{WZ}$ is important in this context since it provides the
source terms for $W(r)$. In the Skyrme model it is found$^{33}$ that the
induced components contribute about $50\%$ to the strange moment of
inertia which demonstrates that these fields may play a crucial role.
Of course, these excitations are also present in the NJL model. A
complete treatment would provide an integral equation for $W(r)$ as the
stationary condition. This, however, is a very tedious task since it
involves complicated matrix elements of $\xi_k$. We therefore
approximate the influence of the induced fields by employing the
derivative expansion. That is, we include all terms up to second
order in the derivatives, especially also the terms which splits
pion and kaon decay constants. The {\it ansatz} (59,60) is then
substituted yielding the strange moment of inertia a functional
of $W(r)$: $\beta^2_I[W]$. This represents a space integral over
linear and quadratic expressions in $W(r)$. The corresponding \
coefficient functions involve the chiral angle for which the
self-consistent static soliton is substituted.

Finally we substitute our ansatz for the rotating meson fields (38)
into the expression for the mesonic part of the action, $\A_m$ (2) and
subtract the $\xi=1$ contribution. This may be done easily and
gives additional contributions to the symmetry breaking parameters
$\gamma,\gamma_1,\gamma_2$ and $\gamma_3$.

As in Skyrme type models Noether charges corresponding to
right $SU(3)$ transformations may be constructed leading to the
quantization prescription for the right $SU(3)$ generators, $R_a$
\be
R_a=-{{\partial L}\over{\partial\Omega_a}}=\cases{
-(\alpha^2\Omega_a+\alpha_1D_{8a})=-J_a,&a=1,2,3\cr
-(\beta^2\Omega_a+\beta_1D_{8a}),&a=4,..,7\cr
\frac{\sqrt3}{2}B,&a=8}
\ee
wherein $J_i\ (i=1,2,3)$ denote the spin operators.

The Hamiltonian operator:
\be
H=-\sum_{a=1}^8R_a\Omega_a-L
\ee
may be diagonalized exactly which is done along the lines of the
original Yabu-Ando approach yielding the energy expression for
baryon $B$:
\be
E_B=E+{1\over2}\big({1\over{\alpha^2}}-{1\over{\beta^2}}\big)J(J+1)
-{3\over{8\beta^2}}+{1\over{2\beta^2}}\epsilon_{SB},
\ee
wherein $\epsilon_{SB}$ is the eigenvalue of
\be
C_2&+&\beta^2\gamma(1-D_{88})
\nonumber \\*
&+&\beta^2({\alpha_1}/{\alpha^2})
\sum_{i=1}^3 D_{8i}(2R_i+\alpha_1D_{8i})+
\beta_1\sum_{\alpha=4}^7D_{8\alpha}(2R_\alpha+
\beta_1D_{8\alpha})\nonumber \\*
&+&\beta^2\gamma_1(1-D_{88}^2)+\beta^2\gamma_2\sum_{i=1}^3 D_{8i}D_{8i}
+\beta^2\gamma_3\sum_{\alpha=4}^7D_{8\alpha}D_{8\alpha} .
\ee
$C_2=\sum_{a=1}^8R_a^2$ denotes the quadratic Casimir operator of
$SU(3)$. The eigenvalues $\epsilon_{SB}$ are determined using
a generalized YA approach\footnote{Yabu and Ando only considered
$C_2+\beta^2\gamma(1-D_{88})$.}  ({\it c.f.} ref.$^{32}$).

The constraint for the right hypercharge
$Y_R=\frac{2}{\sqrt3}R_8=B$
confines the possible eigenstates to those which carry half integer
spin for $B=1$, \ie fermions.

\subsection{Numerical results}

We find that the valence quark dominates the vacuum contribution
to both the non-strange moment of inertia $\alpha^2$ as well as
the strange moment of inertia $\beta^2$. For the latter we also obtain
a significant contribution due to the induced kaon fields. This
result compares with the findings in the Skyrme model.$^{33}$ We would
like to stress that for $M_s\ne M_u$ the vacuum contribution to
the strange moment of inertia $\Theta^{\rm vac}_{44}$ becomes
a non-diverging quantity only after subtracting the reference value
$\Theta^{\rm vac}_{44}(\xi=1)$. Actually we find that the subtracted
value decreases as the difference $M_s- M_u$ increases.

For the coefficients of the expressions
linear in the angular velocities $\Delta_{ab}$ the dominance of the
valence quarks is even more drastically pronounced and the
contribution due to the polarized Dirac sea is almost negligible. For
the relevant range of constituent masses we only find a moderate
dependence of $\Delta_{ab}^{\rm vac}$ on the cut-off $\Lambda$. This
is not surprising since the imaginary part of the Euclidean action
stays finite as $\Lambda\rightarrow\infty$. The linear terms are found
to have a significant influence on the energy eigenvalues of the
baryons. Since both, $\alpha_1$ as well as $\beta_1$ turn out be
negative the linear terms increase the $SU(3)$ symmetry breaking.
In ref.$^{25}$ these linear terms have been incorporated via an expansion
in the angular velocity as well as the strange current mass $m_s^0$
without performing the corresponding shift in the scalar field
({\it c.f.} section 2). Of course, in an evaluation to all orders this
approach should be compatible to ours. However since (at least in the
proper time regularization scheme) we find $M_s-M_u\ll m_s^0$ we
expect the expansion in $M_s-M_u$ to be more reasonable.

In the context of symmetry breaking on the effective Lagrangian level
$\gamma$ - the coefficient of $(1-D_{88})$ - turns out to be most
important. The deviation of $M_s/M_u$
from unity increases $\gamma^m$ by about $25\%$$^{26}$ compared to our
earlier results$^{24}$ where the approximation of identical constituent
masses was made. This increase is compensated by the decrease of
the multiplying integral due to the introduction of the finite pion
mass in the equation of motion for the chiral soliton.
While the contribution of the valence quarks to $\gamma$ is large
and positive the polarized Dirac tends to cancel this effect.

Finally we would like to note that the symmetry breakers involving the
non-diagonal matrix elements of the $SU(3)$ rotations
($\sum_{i=1}^3D_{8i}D_{8i}$ and $\sum_{\alpha=4}^7D_{8\alpha}D_{8\alpha}$)
only play a minor role for the baryon mass differences but nevertheless
they will be included for completeness and consistency.
\begin{table}
\tcaption{The mass differences for the low-lying $\frac{1}{2}^+$ and
$\frac{3}{2}^+$ baryon states as functions of the up-constituent mass
$M_u$. All numbers are in MeV.}
\smallskip
\centerline{
\begin{tabular}{l c c c c c c }
\\
$M_u$ & 350. & 400. & 450. & 500. & Expt. \\
\hline
$M_N$&1684.&1725.&1726.&1717.&938.\\
$M_\Lambda-M_N$&149.&116.&94.&78.&177.\\
$M_\Sigma-M_N$&182.&157.&134.&116.&254.\\
$M_\Xi-M_N$&320.&256.&210&177.&379.\\
$M_\Delta-M_N$&209.&296.&350.&392.&293.\\
$M_{\Sigma^*}-M_N$&350.&401.&433.&460.&446.\\
$M_{\Xi^*}-M_N$&493.&506.&515.&526.&591.\\
$M_\Omega-M_N$&637.&611.&596.&592.&733.\\
\hline
\end{tabular}}
\end{table}

Now we are ready to discuss our predictions for the baryon mass spectrum
which are displayed in table 4. As in almost all types of soliton
models the result for the absolute value of
the masses turns out to be too large when the physical value for $f_\pi$
is used. However, our result is already several hundred MeV lower
than that in the Skyrme model.$^{33,34}$
In the two flavor Skyrme model it has been
demonstrated$^{35,36,37}$ that quantum fluctuations of the classical soliton
significantly lower the absolute energy. It is important to mention that
the dominant subtractions are of $O(N_C^0)$ and therefore identical
for nucleon and $\Delta$. Thus it seems appropriate to concentrate
on mass differences only.

As may been seen from table 4 the mass splittings between states
of different spins decrease with increasing constituent mass. This
is mainly linked to the decrease of $\alpha^2$ with $M_u$. On the
contrary the mass differences between members of the same spin
multiplet get smaller as the constituent mass gets larger since also
$\gamma$ decreases with $M_u$. As mentioned above,
allowing for $m_\pi\ne0$ in the evaluation of the classical soliton
lowers $\alpha^2$ and therefore increases the mass splitting of
the $s=\frac{1}{2}^+$ and $s=\frac{3}{2}^+$ multiplets. Note that
consistency requires to use a non-vanishing current mass, \ie a
non-vanishing pion mass, in the meson as well as in the baryon
sector.

We furthermore recognize form table 4 that the mass splittings between
baryons of the same spin are predicted too low. This shortcoming is
not completely unexpected since already in the meson sector of the model
we have found that \eg for $M_u=400MeV$ $f_K$ is about $15\%$ lower
than the experimental value (cf. table 1). That this deficiency
transfers to the baryon sector may be seen from the following estimate.
We choose $M_u=390MeV$ to reproduce the experimental nucleon-$\Delta$
mass difference. Then we scale the dominant symmetry breaking parameter
$\gamma^m+\gamma^{\rm vac}$ by $(\frac{f_K^{expt.}}{f_K^{pred.}})^2
=\big(\frac{114}{100}\big)^2$ to calculate the baryon mass differences. The
valence quark contribution to $\gamma$ should be described only by
the constituent masses but not by the predicted value for $f_K$.
The results stemming from this estimate are shown in table 5 and
are found to be in excellent agreement with experimental data.

\begin{table}
\tcaption{The mass differences for the estimate described in the
text. The particles' names refer to the corresponding mass difference
with respect to the nucleon.  All numbers are in MeV.}
\smallskip
\centerline{
\begin{tabular}{l c c c c c c c }
\\
$M_u=390 MeV$&$\Lambda$&$\Sigma$&$\Xi$&$\Delta$&$\Sigma^*$&$\Xi^*$&$\Omega$  \\
\hline
calculated&121.&162.&267.&283.&394.&504.&615. \\
$\gamma_{\rm rescaled}$&175.&248.&396.&291.&449.&608.&765. \\
Expt.&177.&254.&379.&293.&446.&591.&733. \\
\hline
\end{tabular}}
\end{table}

Thus we may conclude that the too small predictions for the baryon mass
splittings  are directly connected to the too small result for
$f_K$. We furthermore expect that a more complete model (\ie
(axial-) vector mesons included) which reproduces the experimental
value for $f_K$ will in fact yield agreement for the baryon mass
differences.

\section{Conclusions}

The main conclusions of this talk may be summarized as follows:

\begin{itemize}
\item
The exact evaluation of the quark determinant for the baryon number one
soliton is possible. No derivative or gradient expansion is needed in
order to investigate stability/instability of the soliton.

\item
Restriction of the scalar meson field to its vacuum expectation value
yields stable solitons; without this
restriction the soliton is unstable if the original NJL model is used.
If this model is supplemented with a $\Phi ^4$ term the soliton becomes
stable again. Even more, for a reasonable size of this term it is only
slightly different from the soliton with constraint on the chiral
circle. There are several possible origins for such a term. The
physically best motivated one is the generalization of the NJL model
such that the scale anomaly of QCD is described on an effective level.

\item
The inclusion of the vector meson $\rho $ and the axialvector meson
$a_1$ leads to a valence quark level which has joined the Dirac sea.
Thus {\bf the NJL soliton supports Witten's conjecture, \ie the Skyrmion
picture of the baryon.}

\item
The isoscalar vector meson $\omega $ can only be included if the
imaginary part of the Euclidian action is taken into account. Despite
the fact that the original problem is formulated in Minkowski space
regularization enforces the continuation back and forth to Euclidian space.
The results obtained so far do not indicate that the
above conclusion concerning the Skyrmion picture of the baryon is in
jeopardy. A full calculation including $\pi$, $\omega $, $\rho $ and
$a_1$ is in progress.

\item
The collective approach to the SU(3) chiral soliton gives reasonable
results for the mass splittings of the low--lying $\frac 12 ^+$ and
$\frac 32 ^+$ baryons. The agreement is even excellent after the kaon
decay constant has been rescaled to its physical value.

\end{itemize}

\nonumsection{Acknowledgements}

I would like to thank Prof.\ Holzwarth for his
warm hospitality and the other participants of the workshop
for the stimulating atmosphere and a lot of helpful discussions.

Furthermore, I thank my colleagues H.\ Reinhardt, H.\ Weigel, C.\ Weiss
and U.\ Z\"uckert who participated in parts of the work underlying this
talk. Helpful contributions to this manuscript by H.\ Weigel and U.\
Z\"uckert are gratefully acknowledged.

This work is supported by the Deutsche Forschungsgemeinschaft (DFG)
under contract Re 856/2-1.

\nonumsection{References}

\begin{enumerate}

\item H.\ Reinhardt, these proceedings.

\item Y.\ Nambu and G.\ Jona-Lasinio, {\it Phys.\ Rev.} {\bf 122}
(1961) 345; {\bf 124} (1961) 246.

\item
D.\ Ebert and H.\ Reinhardt, {\it Nucl.\ Phys.} {\bf B271} (1986) 188.

\item
M.\ Schaden, H.\ Reinhardt, P.\ A.\ Amundsen and M.\ J.\
Lavelle, {\it Nucl.\ Phys.} {\bf B339} (1990) 595;

R.\ Alkofer and H.\ Reinhardt, {\it Z.\ Phys.} {\bf A343} (1992) 79.

\item
H.\ Reinhardt and R.\ Alkofer, {\it Phys.\ Lett.} {\bf 207B} (1988) 482;

R.\ Alkofer and H.\ Reinhardt, {\it Z.\ Phys.} {\bf C45} (1989) 245.

\item
V.\ Bernard, R.\ L.\ Jaffe and Ulf-G.\ Mei\ss ner, {\it Nucl.\ Phys.}
{\bf B308} (1988) 753.

V.\ Bernard, Ulf-G. Mei\ss ner, A.\ H.\ Blin and B.\ Hiller,
{\it Phys.\ Lett.} {\bf B253} (1991) 443.

\item
S.\ Klimt, M.\ Lutz, U.\ Vogl and W.\ Weise, {\it Nucl.\ Phys.} {\bf A516}
(1990) 429.

\item
I.\ Aitchinson, C.\ Frazer, E.\ Tador and J.\ Zuk,
{\it Phys.\ Lett.} {\bf B165} (1985) 163.

\item
H.\ Reinhardt and D.\ Ebert, {\it Phys.\ Lett.} {\bf B215} (1986) 459.

\item
H.\ Reinhardt and R.\ W\"unsch, {\it Phys.\ Lett.} {\bf B215} (1988)
577.

\item
T.\ Meissner, F.\ Gr\"ummer and K.\ Goeke, {\it Phys.\ Lett.} {\bf B227}
(1989) 296.

\item
R.\ Alkofer, {\it Phys.\ Lett.} {\bf B236} (1990) 310.

\item
R.\ Alkofer and H.\ Reinhardt, {\it Phys.\ Lett.} {\bf B244} (1991) 461.

\item
R.\ Alkofer, H.\ Reinhardt, H.\ Weigel and U.\ Z\"uckert,
{\it Phys.\ Rev.\ Lett.} {\bf 69} (1992) 1874.

\item
C.\ Sch\"uren, E.\ Ruiz--Arriola and K.\ Goeke, {\it Phys.\ Lett.}
{\bf B287} (1992) 283.

\item
R.\ Alkofer, H.\ Reinhardt, H.\ Weigel and U.\ Z\"uckert,
``The isoscalar vector meson $\omega$ in the Nambu--Jona-Lasinio
soliton", T\"ubingen preprint August 1992; submitted to Phys.\ Lett.\ {\bf
B}.

\item
P.\ Sieber, T.\ Mei\ss ner, F.\ Gr\"ummer and K.\ Goeke, {\it Nucl.\
Phys.} {\bf A547} (1992) 459.

\item
T.\ Mei\ss ner et al., ``Scale Invariance and the Stability of a
Hedgehog Soliton", Bochum preprint RUB-TPII-30/92, Sept.\ 1992.

\item
C.\ Weiss, H.\ Weigel and R.\ Alkofer, ``The Nambu--Jona-Lasinio soliton
with generalized scalar interactions", T\"ubingen preprint, October 1992.

\item
H.\ Reinhardt, {\it Nucl.\ Phys.} {\bf A503} (1989) 825.

\item
K.\ Goeke et al., {\it Phys.\ Lett.} {\bf B256} (1991) 321.

\item
M.\ Wakamatsu and H.\ Yoshiki, {\it Nucl.\ Phys.} {\bf A524} (1991)
561.

\item
H.\ Yabu and K.\ Ando, {\it Nucl.\ Phys.} {\bf B301} (1988) 601.

\item
H.\ Weigel, R.\ Alkofer and H.\ Reinhardt, {\it Phys.\ Lett.}
{\bf B284} (1992) 296.

\item
A.\ Blotz et al., {\it Phys.\ Lett.} {\bf B287} (1992) 29.

\item
H.\ Weigel, R.\ Alkofer and H.\ Reinhardt, ``Strange Baryons as Chiral
Solitons of the Nambu--Jona-Lasinio model", {\it Nucl.\ Phys.} {\bf B},
in press.

\item
J.\ Schwinger, {\it Phys.\ Rev.} {\bf 82} (1951) 664.

\item
U.-G.\ Mei\ss ner, N.\ Kaiser and W.\ Weise, {\it Nucl.\ Phys.} {\bf
A466} (1987) 685.

\item
G.\ Eckart, A.\ Hayashi and G.\ Holzwarth, {\it Nucl.\ Phys.} {\bf A448}
(1986) 732;

B.\ Schwesinger, H.\ Weigel, G.\ Holzwarth and A.\ Hayashi,
{\it Phys.\ Rep.} {\bf 173} (1989) 173.

\item
J.\ Schechter and H.\ Weigel, {\it Phys.\ Lett.} {\bf B261} (1991) 235;
{\it Phys.\ Rev.} {\bf D44} (1991) 2916.

\item
B.\ Schwesinger, H.\ Weigel, {\it Phys.\ Lett.} {\bf B267} (1991) 438;
{\it Nucl.\ Phys.} {\bf A540} (1992) 461.

\item
N.\ W.\ Park and H.\ Weigel, {\it Phys.\ Lett.} {\bf B268} (1991) 155;
{\it Nucl.\ Phys.} {\bf A541} (1992)  453.

\item
H.\ Weigel, J.\ Schechter N.\ W.\ Park and Ulf-G.\ Mei\ss ner,
{\it Phys.\ Rev.} {\bf D43} (1991) 869.

\item
G.\ Pari, B.\ Schwesinger and H.\ Walliser, {\it Phys.\ Lett.}
{\bf B255} (1991) 1.

\item
I.\ Zahed, A.\ Wirzba and Ulf-G.\ Mei\ss ner, {\it Phys.\ Rev.}
{\bf D33} (1986) 830;

A.\ Dobado and J.\ Terron, {\it Phys.\ Lett.} {\bf B247} (1990) 581.

\item
B.\ Moussallam and D.\ Kalafatis, {\it Phys.\ Lett.} {\bf B272} (1991)
196;

B.\ Moussallam, these proceedings.

\item
G.\ Holzwarth, these proceedings.

\end{enumerate}

\end{document}